\documentstyle[epsf,eqsecnum,aps]{revtex}

\tighten
\begin{document}
\twocolumn

\wideabs{
\title{B{\'e}nard-Marangoni Convection in Two Layered Liquids}
\author{Wayne A. Tokaruk\cite{wayne_email},
 T. C. A. Molteno\cite{tim_email}, 
and Stephen W. Morris\cite{smorris_email}}
\address{Department of Physics, University of
Toronto, 60 St. George St., Toronto, Ontario, Canada M5S 1A7 }
\maketitle

\begin{abstract}
We describe experiments on B{\'e}nard-Marangoni convection in horizontal layers of two immiscible liquids. Unlike previous experiments, which used gases as the upper fluid, we find a square planform close to onset which undergoes a secondary bifurcation to rolls at higher temperature differences. The scale of the convection pattern is that of the thinner lower fluid layer for which buoyancy and surface tension forces are comparable.  The wavenumber of the pattern near onset agrees with the prediction of the linear stability analysis for the full two-layer problem.  The square planform is in qualitative agreement with recent one- and two-layer nonlinear theories, which fail however to predict the transition to rolls.\\ \\
Submitted to {\it Physical Review Letters}, June 29, 1998. See also http://mobydick.physics.utoronto.ca.

\end{abstract}
}

\narrowtext

Convection in fluids has been a fruitful system for the study of 
nonlinear, nonequilibrium patterns for almost 100 years.\cite{benard,normand,ch93} The original experiments of B{\'e}nard\cite{benard} used shallow layers 
of whale oil, heated from below and open to the air above.  Many years passed before it was conclusively shown that surface tension gradients, or ``Marangoni'', forces were crucial.\cite{block,pearson} In general, both Marangoni and buoyancy forces are present.\cite{nield,nepobook} Beginning with Pearson\cite{pearson}, theories of this instability have traditionally neglected the dynamics of the upper fluid (air in B{\'e}nard's experiments), replacing it by a constant heat flux boundary condition on the lower fluid.  This condition is experimentally unrealizable, however.  In well-controlled experiments\cite{kosch_big,kosch,schatzPRL,thess,vanhookPRL,vanhookJFM}, the upper fluid is bounded above by a plate, on which a constant temperature boundary condition is maintained. The heat flux across the interface is then the result of conduction and convection in both fluids, and is not constant above the onset of convection. The two-fluid interface is deformable, and its distance from the upper plate also changes the local heat flux.\cite{vanhookJFM}  In addition, buoyancy forces in the upper fluid may actively assist or impede convection in the lower fluid {\it via} surface stresses.\cite{nepobook,zeren,ferm} Recent experiments have found it necessary to treat the dynamics of both fluids in order the quantitatively explain the data.\cite{vanhookJFM} 

In order to gain insight into the more complex, but experimentally better posed, two-layer problem, we have undertaken an experimental study of convection in a system of two immiscible fluids sharing a deformable interface. To our knowledge, only one previous experiment has been done on this system, and it was restricted to locating the onset of convection, without visualization\cite{zeren}.  On the theoretical side, the two-layer problem has been the subject of a complete linear stability analysis\cite{nepobook,zeren,ferm} and some weakly nonlinear analyses\cite{nepobook,golovin} in certain limiting cases.  The general problem, including surface deformations, should however be well within the range of modern weakly nonlinear theory and of numerical simulations.     

In our experiment, both fluids were liquids, and the conditions were such that the Marangoni forces were comparable to the buoyancy forces at onset.  The dimensionless heat flux across the interface is larger in our experiment than in previous studies. This is expected to have important effects in the nonlinear regime.\cite{regnier,golovin}  We used shadowgraph imaging to visualize the pattern and studied its secondary instabilities.  We measured the mean wavenumber of the patterns, and compared them to the linear stability result, calculated for the full two-layer problem\cite{ferm}.  We could not resolve any hysteresis at the onset of convection.  The first clear pattern we found above onset had a square planform.  The mean wavenumber of this pattern is in good agreement with the critical value predicted from linear theory\cite{ferm}.  The square planform persisted up to $\epsilon \sim 0.7$, where it underwent a slightly hysteretic transition to a roll pattern.  Here, $\epsilon = (\Delta T/\Delta T_c)-1$, where $\Delta T$ is the temperature difference across {\it both} layers, and $\Delta T_c$ is its value at the onset of convection. Up to the maximum value of $\epsilon \sim 1.4$ we reached, the scale of the convection cells was determined by the depth of the lower liquid. We compare our results with two recent nonlinear theories\cite{regnier,golovin} and find qualitative agreement.   
   
The lower fluid was FC-75, a low-viscosity, perfluorinated hydrocarbon\cite{3Minfo}, while the upper was water.  An important parameter is the depth fraction,  $L = d^+ / d$, where $d^+$~($d$) is the depth of the upper~(lower) fluid. In this paper, we report results for $L = 2.18 \pm 0.04$. The dimensionless heat flux across the interface, which is uniform below the onset of convection, is given by the Biot number ${\cal B} = \Lambda^+ / \Lambda L$, where $\Lambda^+$~($\Lambda$) is the thermal conductivity of the upper~(lower) fluid\cite{biotfoot}. In contrast to previous liquid/gas experiments\cite{kosch_big,kosch,schatzPRL,thess,vanhookPRL,vanhookJFM}, for which ${\cal B} \sim 0.1$, in our experiment $\Lambda^+ > \Lambda$ so that most of the total temperature drop $\Delta T$ falls across the thinner lower layer and ${\cal B} = 4.31 \pm 0.08$ at onset.   

Water, at a constant temperature $T_0$, bathed the top surface of a sapphire window which formed the top boundary of the cell.  The base plate of the cell was 
\begin{figure}
\epsfxsize=3.7in
\centerline{\epsffile{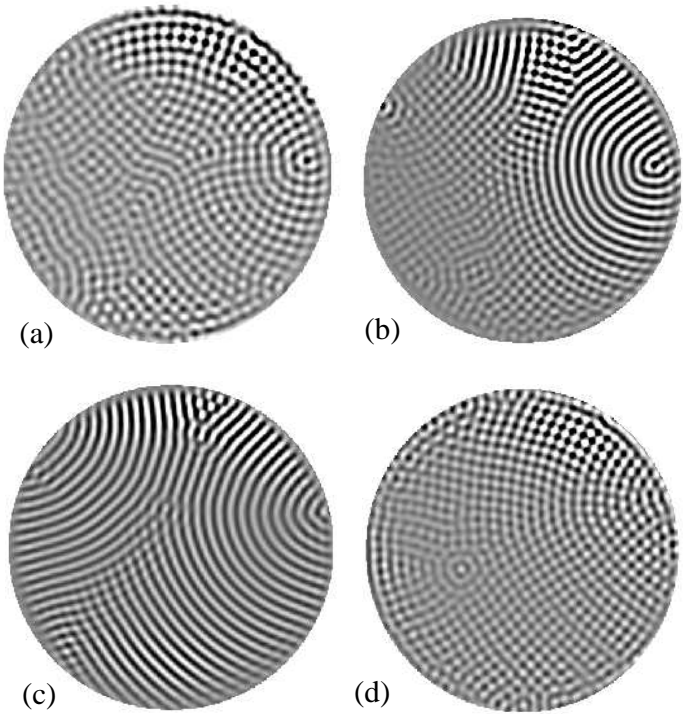}}
\vskip 0.1in
\caption{Shadowgraph images of the patterns observed.  (a) A disordered square planform at $\epsilon = 0.48$.  (b) Near the transition from squares to rolls at $\epsilon = 0.78$ , coexistence is observed. (c) Roll planform at $\epsilon = 1.38$. (d) A more ordered square pattern observed upon slowly decreasing $\epsilon$ to 0.50, starting from roll pattern (c). }
\label{pictures}
\end{figure}
\vskip 0.1in
maintained at a temperature $T_1>T_0$ using a thin-film electric heater attached to its bottom surface.  Both temperatures were feedback controlled to $\pm 1$mK. $\Delta T = T_1 - T_0$ is the experimental control parameter.  We estimated the heat flux through the cell by monitoring the heater power.  The sidewall of the cell was made of plastic and included a wedge-shaped radial fin to stabilize the two-fluid contact line.  The round cell had thickness $D = d^+ + d = 4.06 \pm 0.01$ mm, and radius $r=41.15 \pm 0.05$ mm.  The relevant aspect ratio is that of the lower fluid, $\Gamma = r/d = 32.3$. The uniformity of $D$ was $\pm 8 \mu$m, checked interferometrically.  The cell was leveled to within $\pm 100 \mu$rad, using an electronic bubble level.

Both fluids wet all of the surfaces, so bridged configurations of the interface were difficult to avoid.  To prepare the two-layer configuration, the horizontal cell was completely filled with FC-75 which was then partially displaced by water slowly introduced above the fin.  The interface was made level with the fin by visually eliminating the meniscus. The resulting metastable layering was somewhat imperfect due to the pinning of small deformations of the contact line, and because the interface was difficult to see. The filling fraction $L$ was determined destructively at the end of the experiment by tipping the cell on edge and measuring the interface position.  

The onset of convection and the nonlinear patterns were visualized by the shadowgraph method\cite{rsi}, using a beam reflected off the mirrored bottom plate of the cell.  This beam passed twice through each fluid and twice through the interface.  The shadowgraph signal therefore contained contrast due to temperature gradients in both fluids, plus a component due to deflections at the deformed interface.  The latter effect is likely to be small, since the two fluids are nearly index matched.  We observed that the scale of the pattern in the upper layer was slaved to that of the lower.  Thus, it was not necessary to disentangle the various contributions to the shadowgraph images in order to understand the basic planforms.   

A typical run consisted of slowly increasing and decreasing $\Delta T$, while recording images and measuring the heat flux. The onset of convection occurred at $\Delta T_c = 0.999 \pm 0.025^\circ C$.  Convection entered from one side of cell as a highly disordered pattern, and eventually filled the cell.  We did not observe hysteresis in the onset, presumably because of the rounding effect due to nonuniformities in the cell.  The first clear patterns that emerged had a square planform.  Square patterns have been observed previously in Marangoni convection\cite{thess}, but only at much higher $\epsilon$ as a secondary bifurcation from a hexagonal pattern.  Here, our observations are consistent with a square planform emerging at onset, but, due to rounding effects, we cannot exclude the possibility that a hexagonal pattern exists in a narrow range close to onset.

Typical patterns are shown in Figure \ref{pictures}. For increasing $\epsilon$, a patchy, time-dependent square planform, as shown in Figure \ref{pictures}(a) was found between onset and $\epsilon_{sr} = 0.70 \pm 0.01$ where a slightly hysteretic transition to rolls was observed. This transition could easily be located by a significant increase in the heat flux through the cell. Near the transition, we find a dynamical coexistence of squares and rolls, as shown in Figure \ref{pictures}(b). The rolls appeared where squares merged along shared edges, and {\it visa versa}.  At values of $\epsilon > \epsilon_{sr}$, the pattern was dominated by rolls, shown in Figure \ref{pictures}(c), with squares appearing only at grain boundaries.  The more ordered roll patterns were very slowly time dependent. The finned sidewalls had a rather weak orienting effect on the rolls.  As $\epsilon$ was decreased, an ordered square pattern re-emerged at $\epsilon_{rs} = 0.67 \pm 0.02$ from the pre-existing rolls.  The hysteresis $\epsilon_{sr} - \epsilon_{rs} = 0.03$ was just barely resolvable from the heat flux data.  Figure \ref{pictures}(d) shows such a square planform.  In all cases, the patterns had a scale $\sim d$, the dimension of the thinner lower layer.  

The characteristic timescale for convection in the lower fluid is the vertical thermal diffusion time $\tau_v = d^2/\kappa = 47$s, where $\kappa$ is the thermal diffusivity.  We ramped the temperature slowly compared to $\tau_v$, but it was impractical to increase it more slowly than the much-longer horizontal diffusion time $\tau_h = \Gamma^2 \tau_v = 14$hrs.  Thus, some of the disorder we observe on increasing $\epsilon$ may be due to the finite ramp rate.  However, we also performed experiments in which $\epsilon$ was ramped very quickly to a value $< \epsilon_{sr}$, and then held constant for $\sim 6 \tau_h$.  The resulting 
\begin{figure}
\epsfxsize=3.5in
\centerline{\epsffile{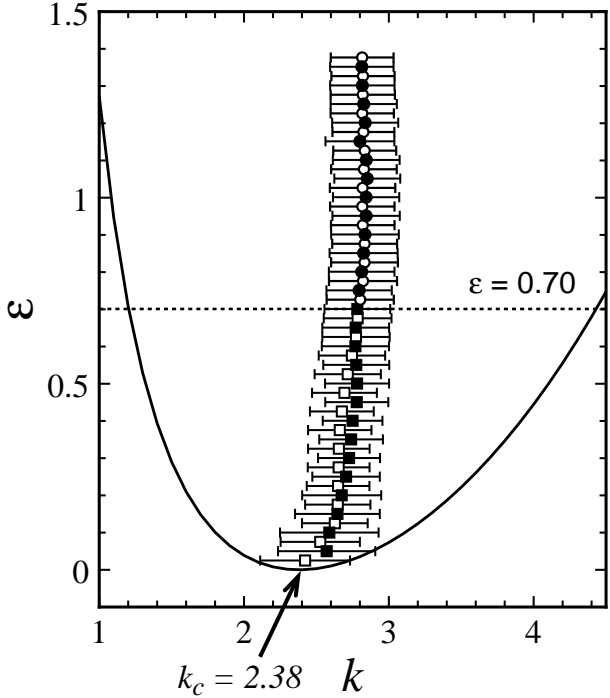}}
\vskip 0.1in
\caption{The mean wavenumber $\langle k \rangle$, showing the linear stability boundary from Ref.~\protect{\cite{ferm}} (solid line). Bars indicate the spread in $\langle k \rangle$ due to disorder in the pattern. We find a square planform (square symbols) for $\epsilon < 0.7$, and rolls (circular symbols) otherwise.  Open (solid) symbols indicate increasing (decreasing) $\epsilon$. }
\label{wavenumber}
\end{figure}
\vskip 0.1in
square pattern did not anneal significantly and remained time-dependent. Thus, the disorder may be partially dynamical in origin.     

We determined the wavenumber of the patterns from Fourier analysis.  We found the azimuthally averaged structure function of a $256^2$ pixel region of the centre of the cell.  The peak in the structure function gave the mean wavenumber $\langle k \rangle$.  Fig.~\ref{wavenumber} compares $\langle k \rangle$ with the results of linear stability theory\cite{ferm}. The dimensionless wavenumber $k$ is scaled by the depth of the bottom fluid $d$.  The value of $\langle k \rangle$ at small $\epsilon$ agrees with the critical value $k_c = 2.38$ from the linear theory. The more ordered square patterns observed for decreasing $\epsilon$ have slightly larger $\langle k \rangle$ than those of increasing $\epsilon$. 

To evaluate the linear stability boundary shown in Fig.~\ref{wavenumber}, all of the material parameters are needed.  These are, for the lower fluid, the kinematic viscosity  $\nu$, the density $\rho$, and $\kappa = \Lambda/\rho C_P$, where $C_P$ is the heat capacity, as well as the corresponding quantities for the upper fluid and the interfacial surface tension $\sigma$.  All of these are assumed to be independent of $T$, according to the usual Boussinesq approximation, except $\sigma$ and the two densities, which are linearly temperature dependent.  Thus, we also require the thermal expansion coefficient $\beta =(1/\rho)\partial\rho/\partial T$ for each fluid and the temperature derivative of the interfacial surface tension $\gamma = - \partial\sigma/\partial T$.  The negative sign is included so that $\gamma > 0$. All of the parameters are known, except $\sigma$ and $\gamma$.  Using Antonow's rule\cite{ant}, we estimated $\sigma$ to be equal to the difference of the surface tensions of the two fluids, measured against air.  This approximation suffices, because the neutral stability boundary is very insensitive to $\sigma$, which only enters into the small surface deformations\cite{prandtl_etc}.  On the other hand, the remaining parameter $\gamma$ is crucial to the Marangoni mechanism.  We fixed $\gamma$ by requiring that the linear theory reproduce the correct measured $\Delta T_c$.  The result is $\gamma = 0.047 \pm 0.003$ dynes/cm K.  This is consistent to within a factor of two with the difference of the known values of $\gamma$ for each fluid, measured against air (i.e. with Antonow's rule\cite{ant} applied to $\partial\sigma/\partial T$).

One- and two-layer theories of B{\'e}nard-Marangoni convection traditionally scale the problem using the temperature difference $\Delta T^L$ across the {\it lower} fluid.  Below onset,
\begin{equation}
\Delta T^L = (1+\Lambda L/\Lambda^+)^{-1}\Delta T = (1+1/{\cal B})^{-1}\Delta T .
\label{DTlower}
\end{equation}
This scaling is only approximate above onset, where convection effectively increases the average thermal conductivity of each layer, and also introduces temperature variations on the interface\cite{nepobook,perezgarcia98}.  The often-used assumption of a constant ${\cal B}$ is problematical above onset for the same reasons.\cite{biotfoot,perezgarcia98}  However, in order to make contact with theories using this scaling, it is customary to adopt Eqn.~\ref{DTlower} even above onset.\cite{kosch_big}  The seriousness of this approximation can be appreciated from heat flux measurements.\cite{perezgarcia98}  The total effective thermal conductivity of the cell at the highest $\epsilon$ was only about 8\% larger than that below onset.  This sets an upper bound on the validity of Eqn.~\ref{DTlower}.

Assuming Eqn.~\ref{DTlower}, one can define the Rayleigh and Marangoni numbers ${\cal R}$ and ${\cal M}$ of the lower layer as follows;
\begin{equation}
{\cal R} = \bigg[\frac{g \beta d^3}{\kappa \nu}\bigg] \Delta T^L {~~~~~~~}
{\cal M} = \bigg[\frac{d \gamma}{\rho \kappa \nu } \bigg] \Delta T^L \,,
\label{MRnumber}
\end{equation}
where $g$ is the acceleration due to gravity.  ${\cal R}$ and ${\cal M}$ are not independent, so it is convenient to define two related parameters \cite{parmentier},
\begin{equation}
\alpha = \bigg[1+\frac{M}{M_0}~\frac{R_0}{R}\bigg]^{-1}
{~~~~~~}
\lambda = \frac{R}{R_0}+\frac{M}{M_0}\,.
\label{alpha lambda}
\end{equation}
Here, ${\cal R}_0$ (${\cal M}_0$) is the critical value of ${\cal R}$ (${\cal M}$) in the absence of Marangoni (buoyancy) forces.  In general, in the two-layer problem ${\cal R}_0$ and ${\cal M}_0$ will differ from the usual critical values for any one-layer model of the lower fluid. They can be determined from the full linear theory of the coupled, two-layer system, as described below.  $\alpha$ is fixed by the choice of fluids and the depth $d$, and can be considered as the fraction of the total forcing due to buoyancy, while $\lambda$ is proportional to the total temperature difference $\Delta T$.
\begin{figure}
\epsfxsize=3.5in
\centerline{\epsffile{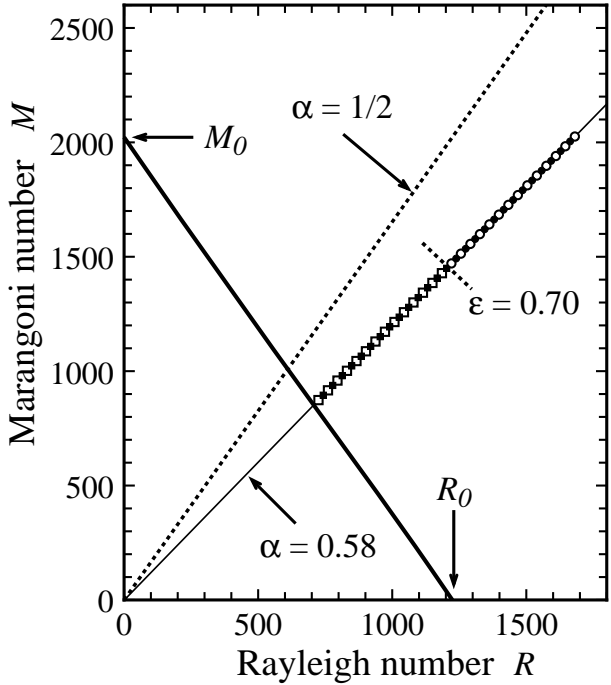}}
\vskip 0.1in
\caption{The Marangoni number ${\cal M}$ {\it vs.} the Rayleigh number ${\cal R}$ for the lower fluid, showing the neutral stability boundary (thick solid line) from the complete two-layer analysis of Ref.~\protect{\cite{ferm}}.  Open (solid) symbols show data for increasing (decreasing) $\epsilon$.  Square (circular) symbols represent points where the square (roll) planform was observed. }
\label{M_vs_R}
\end{figure}
\vskip 0.1in
Fig.~\ref{M_vs_R} shows a plot of ${\cal M}$ {\it vs.} ${\cal R}$.  The linear stability boundary for the coupled, two-layer system at constant $L$ separates the conduction regime near the origin from the convecting regime.  Its intercepts are the critical values ${\cal R}_0$ and ${\cal M}_0$ of each number in the absence of the complementary effect.  The boundary is calculated by holding $L$ and the material parameters constant and varying $d$.  One point on this line corresponds to the experimental value of $d$. $\alpha$ varies from 0 on the ${\cal M}$ axis to 1 on the ${\cal R}$ axis. Above the line $\alpha = 1/ 2$, Marangoni forces dominate; below, buoyancy.  For our experimental conditions, $\alpha = 0.58$, so that the two effects are nearly equal. Assuming Eqn.~\ref{DTlower} holds, increasing $\lambda \propto \epsilon$ corresponds to moving outward along the radial line $\alpha = 0.58$.  The symbols show the values of ${\cal M}$ and ${\cal R}$ that we reached, and the point where the planform changed from squares to rolls. It would be interesting to map the stability boundaries of various patterns in the $({\cal R},{\cal M})$ plane by experimentally varying $\alpha$ and $\lambda$ independently.  This can be done in principle by varying $\epsilon$ and $d$, keeping $L$ fixed.

Two recent nonlinear theories are relevant to our observations\cite{regnier,golovin}, although neither treats the full two-layer problem.  Regnier {\it et. al.}\cite{regnier} included both surface tension and buoyancy in a one-layer model with no surface deformations and a constant ${\cal B}$ boundary condition on the upper surface.  They derived a set of amplitude equations which show various regimes of hexagonal, square and roll planforms above onset. Their main result is that there exists a critical value $\alpha_c$ above which a hexagonal pattern at onset gives way to squares for $\alpha < \alpha_c$ or to rolls for $\alpha > \alpha_c$. Unfortunately, they only present results for ${\cal B} < 2$, where our ${\cal B} \sim 4$. The trends however suggest that as ${\cal B}$ is increased in their model, $\alpha_c \rightarrow 1$, favouring squares. In the same limit, the range above onset over which the hexagonal pattern exists becomes very narrow.  These two predictions are in qualitative agreement with our observations at low $\epsilon$.  

Golovin {\it et. al.}\cite{golovin} treated the full two-layer problem, but included Marangoni forces only.  They derived amplitude equations, including the effect of surface deformations, but neglecting stresses on the interface on the assumption that the upper fluid was a gas. Interestingly, when the heat transfer through the upper fluid (i.e. ${\cal B}$) is increased, a square pattern is predicted at onset.  This is also consistent with our observations.  Neither of these nonlinear theories predicts the square to roll transition we observed, however.  Nor do the amplitude equations contain the gradient terms that would be necessary to describe disordered patterns of the kind we observe. 

In summary, we have performed experiments on thermal convection in a system of two layered liquids under conditions where both buoyancy and Marangoni forces are comparable.  We found a square planform of convection just above onset, which underwent a transition to a roll pattern at higher control parameter.  We compared the wavenumber of the patterns to the linear stability boundary for the full two-layer coupled problem. We found quantitative agreement between the wavenumber of the pattern just above onset and the critical wavenumber $k_c$ from linear theory.  Two nonlinear theories, originally proposed to explain gas over liquid experiments, can be applied to our case. They qualitatively predict the square patterns found near onset, but not the observed transition to rolls.  

We thank Stephen VanHook, Harry Swinney and Zahir Daya for useful discussions. This research was supported by the Natural Sciences and Engineering Research Council of Canada and the Walter C. Sumner Memorial Fellowship.


\begin{references}
\bibitem{3Minfo} FC-75 is manufactured by 3M Inc.  The saturation concentration of water in FC-75 is only 11 ppm, so the assumption of immiscibility should be well satisfied.
\bibitem{biotfoot} In a one-layer model with a passive upper layer, the Biot number just above the onset of convection is given by ${\cal B}(k) = \Lambda^+ k / \Lambda \tanh( k L )$, where $k$ is the dimensionless wavenumber of the pattern (See Refs.~\protect{\cite{normand}} and \protect{\cite{perezgarcia98}}).  In a full two-layer model, the heat transfer is better described by the two-layer Biot number ${\cal F} = (L^{-1}-{\cal B})/(1+{\cal B})$ (See Ref.~\protect{\cite{vanhookJFM}}).
\bibitem{prandtl_etc} From the material parameters, one can define several useful dimensionless numbers. The small Crispation number ${\cal C} = \rho \nu \kappa/\sigma d \sim 8 \times 10^{-7}$, implies that surface deformations are probably negligible.   The Galileo number $G = g d^3 / \nu \kappa$ is $\sim 6 \times 10^5$, so that we do not expect to observe long wavelength instabilities (See refs.~\protect{\cite{vanhookPRL}} and \cite{vanhookJFM}). Finally, the Prandtl number ${\cal P} = \nu/\kappa$ is the ratio of the thermal and viscous relaxation times.  For our fluids, ${\cal P}^+ \approx 7$ and ${\cal P} \approx 27$, so that inertial effects are small.
\end{references}
\end{document}